# NINETEENTH CENTURY AIR POLLUTION VARIATIONS IN PARIS INFERRED FROM EIFFEL TOWER POTENTIAL GRADIENT MEASUREMENTS

R.G.Harrison[1] and K.L. Aplin[2]

[1]Department of Meteorology, The University of Reading, P.O Box 243, Earley Gate, Reading Berks, RG6 6BB UK
[2]Space Science and Technology Department, Rutherford Appleton Laboratory, Chilton, Didcot, Oxon, OX11 0QX UK

ABSTRACT Early surface measurements of atmospheric Potential Gradient were made in many European cities in the nineteenth century (C19th). The data was usually obtained at hourly resolution, and good accounts of the calibration of the instruments are also often available. The PG measurements made by Chauveau on the Eiffel Tower, soon after its completion in the 1890s, are particularly notable. Atmospheric electrical proxy techniques in combination with simple boundary layer meteorology are used to determine air pollution levels. The C19th PG measurements in both polluted and clean Parisian air present a unique resource for European air pollution and atmospheric composition studies.

## INTRODUCTION

Measurements of the Potential Gradient (PG) have been made over a long period, mostly at the surface and frequently in urban centres. Aerosol pollution reduces the electric conductivity of atmospheric air, and increases the PG. In air in which the PG variations are dominated by aerosol variations, the well-established relationship between PG and aerosol permits the PG to be used as a proxy from which the aerosol variations have been inferred [*Harrison and Aplin*, 2002] near London for 1862-1864. Historical atmospheric electrical data from Paris is considered here, from which smoke pollution in the C19th is estimated using a similar technique. Reconstruction of past urban pollution is one aspect of understanding changes in atmospheric composition arising from anthropogenic activity.

Hourly observations were made in Paris in the C19th, both at the *Bureau Central Meteorologique* and, in 1893 at the top of the Eiffel Tower (ET) [*Chauveau*, 1925]. The measurements of potential were made using a fibre electrometer, and are thoroughly described in *Chauveau* [1925]. The ET measurements were made in summer and autumn, and, as for the surface measurements, were presented as absolute values.

## DAILY VARIATIONS

The diurnal variation in PG in clean oceanic air is has been established to follow a unitary variation, with a maximum at 19UT and a minimum at 03UT. This global atmospheric electrical diurnal variation was identified in data obtained on voyages of the geophysical research vessel, *Carnegie*, between 1915 and 1929, and a similar form has been found in clean air throughout the C20th. Hourly PG data from polluted cities shows a very different diurnal variation, frequently with an additional maximum in the morning. At Kew, two daily maxima were recorded through the period of measurements from 1898-1931 [*Scrase,* 1934]. A small change in the phase of the maxima after the introduction of summer time in 1915 [*Scrase,* 1934], established that the changes were of local origin. Direct smoke pollution measurements subsequently begun at the same site showed a diurnal variation with a similar double maximum behaviour. Such variations are a typical signature of polluted air, as they show when sources of particles are usually at their greatest, either from vehicular sources, or, before the advent of motor traffic, domestic heating. The close similarity between PG and simultaneous direct smoke measurements at Kew permits a semi-empirical calibration [*Harrison and Aplin*, 2002].

Figure 1 shows the daily variation in PG measured by Chauveau in 1893. The surface measurements show a double maximum behaviour. At the top of the ET (324m high), the PG shows only a single maximum. (The PG at the top of the ET is much greater than the surface value, because the tower is an earthed structure.) The amplitude of the maximum to minimum variation is, however 43%, which is greater than the typical Carnegie variation, indicating that the diurnal variation results from a combination of global and local effects. Local perturbations at both the measurement sites are very likely to be caused by smoke from pollution.

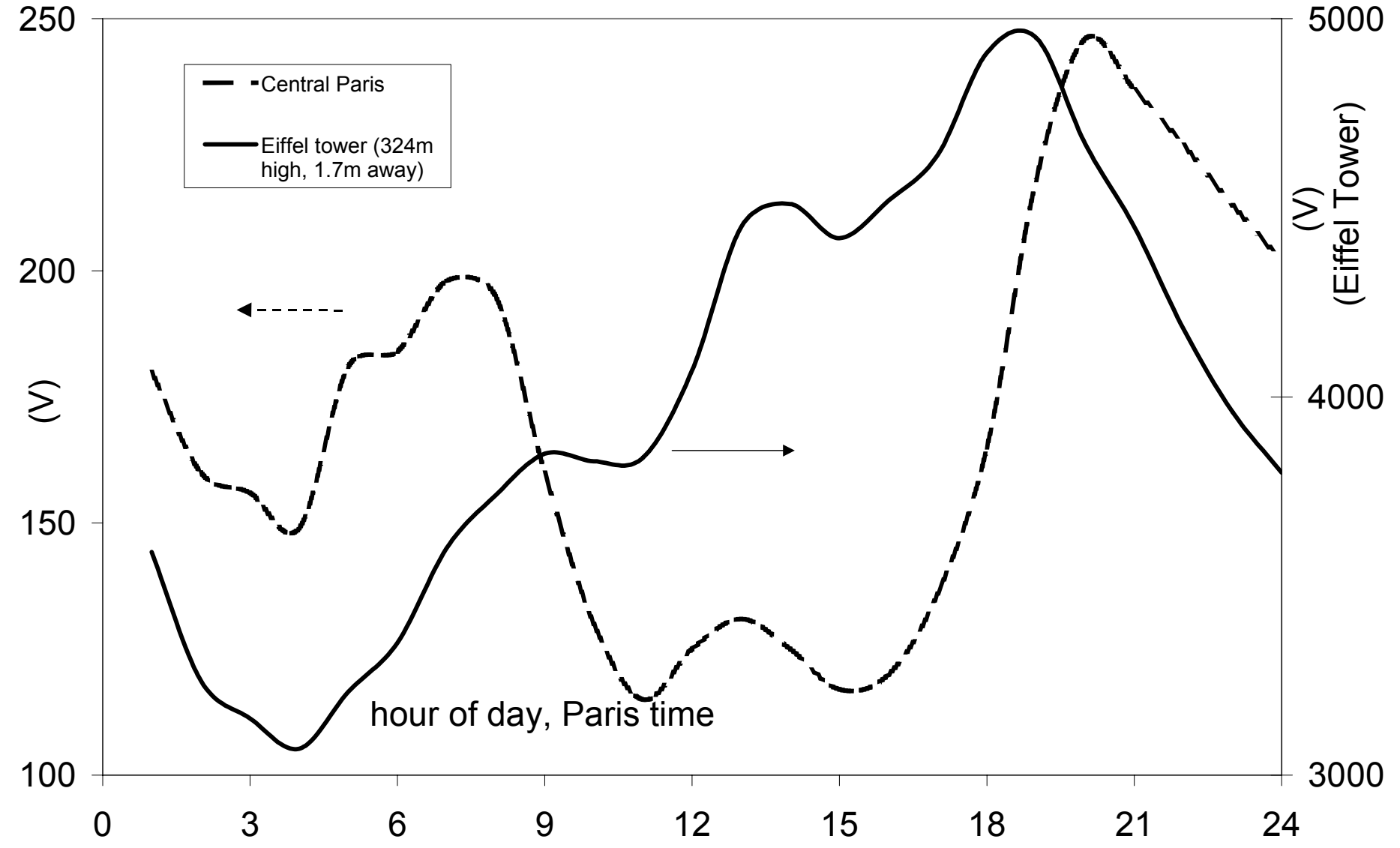

**Figure 1 Diurnal variation of Potential at the surface in Central Paris, and at the top of the Eiffel Tower.**

METHODOLOGY

The ET electrical data present a unique set of measurements from which daily variations in C19th Parisian smoke pollution can be studied. An understanding of other local factors is, however, also required. During the day, the urban boundary layer (BL) varies in depth. The nocturnal BL is shallower than the height of the tower, and the growing daytime convective BL engulfs the tower as it grows [*Dupont et al*, 1999]. There are therefore two different regions of air from to which the summit of the ET is exposed, free tropospheric air and urban BL air. Making the assumption that the free tropospheric air is clean, the electrical variations can be expected to be dominated by global atmospheric electrical changes, as given by the Carnegie curve. The variations found at other times result from the additional effect of smoke pollution. A BL model is used to estimate the typical diurnal variation in BL height, compared with direct data obtained recently [*Dupont et al*, 1999]. Based on this model, the C19th ET measurements likely to have been obtained in free tropospheric and BL air are treated separately.

BOUNDARY LAYER MODEL DESCRIPTION

*Carson's* [1973] parameterisation for growth of the daytime BL from convective turbulence has been used. This approach assumes that a stable temperature gradient $\gamma$ arises from nocturnal cooling, and the BL begins to grow to a height $H$ as soon as the sun rises. The height of the daytime BL is proportional to the total heating received by the BL $S_g$ and is given by:

$$H^2 = \frac{2(1+2A)}{\rho C_p \gamma}\int_0^t S_g dt \quad (1)$$

where $\rho$ is the density of air, $C_p$ the heat capacity of air at constant pressure, $t$ time in seconds and A an empirical dimensionless constant (0.01). The total heating was calculated by estimating the direct solar radiation for Paris (48.7ºN, 2.4ºE) using the method described in *Aplin and Harrison* [2002] for a typical equinoctal day length of 12 hours. The rate of growth of the nocturnal BL *dH/dt* is given by *Garratt* [1994] as $dH/dt \approx \sqrt{0.2\kappa u_* L t}$ . Typical values of the von Karman constant $\kappa$=0.4, friction velocity $u_*$=0.1 ms$^{-1}$, and the length scale of turbulence L=2m were used to estimate a growth rate of 12 m hr$^{-1}$. Modern measurements of the BL height made over four days in March [*Dupont et al*, 1999] were used to constrain the model estimates. *Carson's* [1973] simple model assumes that the BL continues to grow until sunset, which is inconsistent with observations showing that the BL decayed from ~1600UT. An empirical linear scaling for the transition into the nocturnal BL was determined from the rate of decay of the daytime BL observed by *Dupont et al* [1999]. This determines the BL height until 04UT, when the growth of the nocturnal BL exceeds the decaying convective BL.

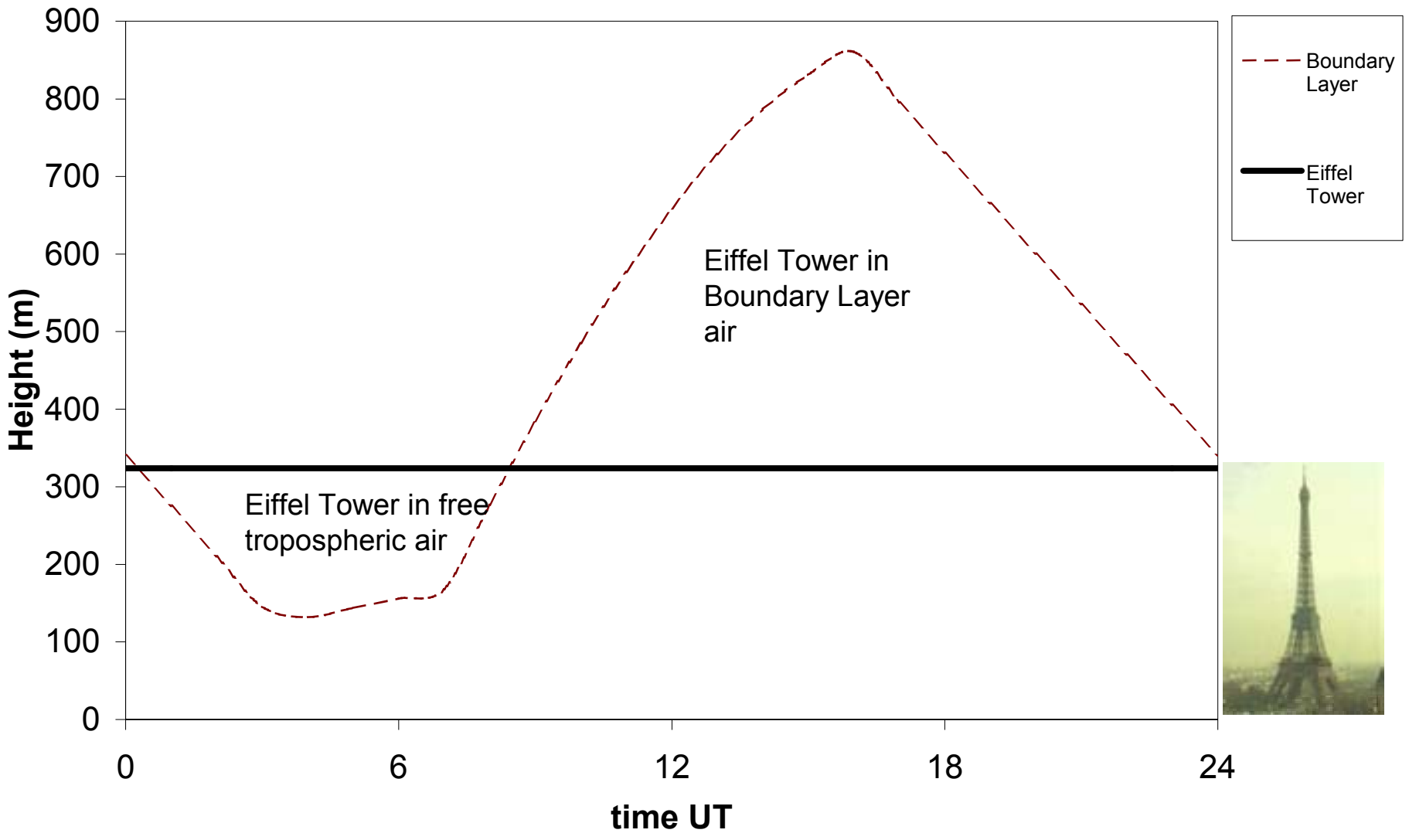

**Figure 2 Evolution of the Boundary Layer for a day length of 12 hours. The Eiffel Tower summit height is indicated as a solid line.**

At sunrise (06UT) the convective BL begins to grow, reaching its maximum height at 16UT, Figure 2. Turbulent eddies decay in the absence of solar heating until 03UT when the nocturnal BL dominates until soon after sunrise. The top of the ET is in free tropospheric air from ~00-09UT. For the rest of the day, the ET is within the polluted air of the growing or decaying convective BL.

During the 00-09UT period when the ET is in the free troposphere, it is assumed that the ET variations are from the clean air Carnegie variation. It is possible to use the relationship between the Carnegie variation and the ET PG for these times to calibrate the ET PG to standard surface oceanic values. The Carnegie values used are the annual values [*Israel*, 1970] and not seasonally adjusted. Minima were fitted to estimate Parisian local time as equivalent to UT+1, as it is today. A linear regression between the relative variation in ET PG and the absolute Carnegie PG values (C), for the times when the ET is in the free troposphere, was used to estimate the ET PG, corrected for the height of the grounded tower. The regression equation is

$$ETPG = (0.5 \pm 0.1)C - (26.7 \pm 13.1) \quad (2).$$

Figure 3 shows the PG at the top of the ET, calculated from (2). The ET PG shows an afternoon peak a few hours before the classical Carnegie maximum, but declines at the same rate as the Carnegie PG in the evening. The PG for the times when the top of the ET is in the free troposphere closely matches the Carnegie variation. A polluted atmosphere is expected to show a larger PG than that from the Carnegie variation. The early evening ET PG onwards is similar in shape to the Carnegie PG, but it is unclear whether the data from 1800 is dominated by local or global effects. However, the excess in ET PG compared to the Carnegie from 09-17UT, when the ET is within the polluted BL, is very likely to be due to smoke pollution.

The sensitivity of the PG to smoke was found at Kew in the early twentieth century by *Harrison and Aplin* [2002], and similar combustion product sizes are assumed for nineteenth century Paris. The excess of the ET PG over the Carnegie PG was used to find the smoke pollution levels for 09-17UT. Surface electrical measurements made at the *Bureau Central Meteorologique* were calibrated in the same way, using the morning PG maximum, which can only have arisen from local effects.

On this basis, the morning maximum Paris surface pollution is 37±10μgm$^{-3}$, much lower than the Kew value of 170±50μgm$^{-3}$ calculated by *Harrison and Aplin* [2002]. The midday smoke concentration at the top of the ET is still lower, 20±80 μgm$^{-3}$. The errors in calculating the smoke concentration above the surface are greater because two regressions are used, firstly to calibrate the absolute value of PG, and secondly to calibrate smoke to PG, with cumulative errors. As air within the convective BL is usually well-mixed, the pollution at the top of the ET is comparable with that at the top of the BL from 09-17UT. On this assumption, it is also possible to infer a profile of pollution based on the BL height, from 09-17UT, Figure 4.

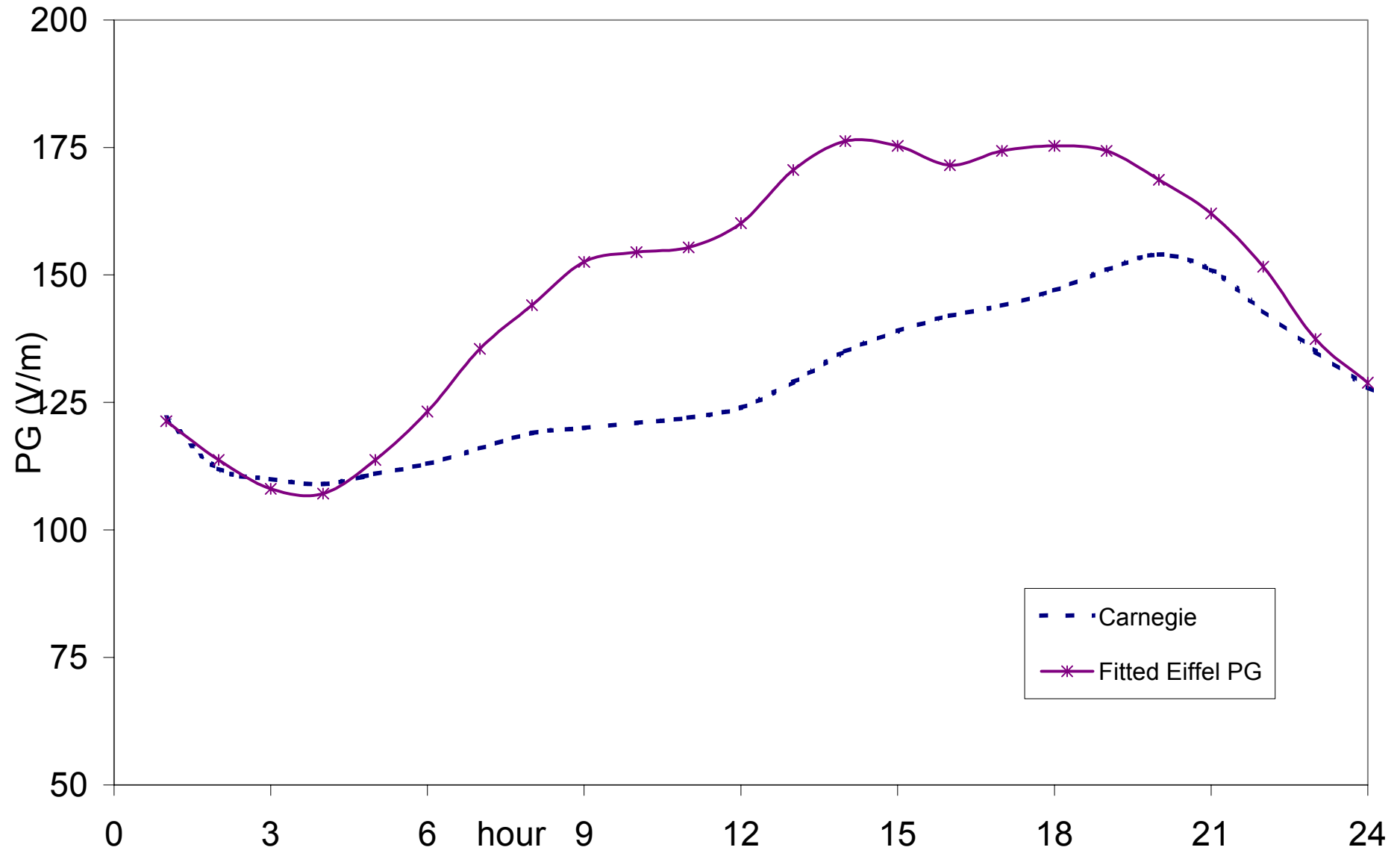

**Figure 3 Estimated Potential Gradient at the top of the Eiffel Tower compared to the clean air Carnegie variation.**

DISCUSSION

The mean smoke pollution in 1893 Paris is substantially lower than the values for Kew (London) in 1863, and the Parisian PG is not dominated by smoke to the same extent. However the C19th Parisian PG variation has a significant local component, both at the surface and aloft. This work also illustrates that the presence of a unimodal PG variation is a necessary, but not sufficient criterion from which to conclude that measurements are globally representative.

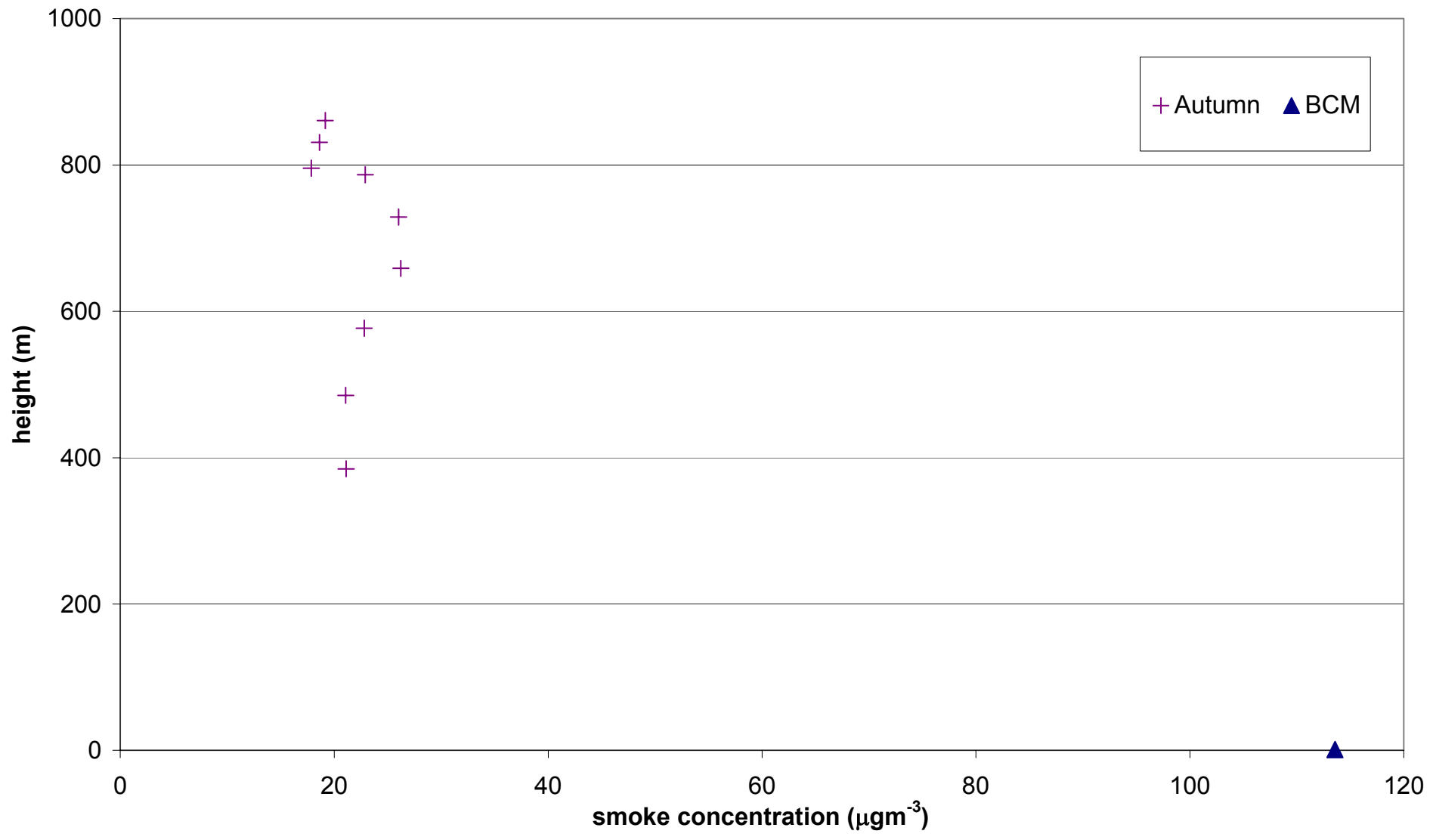

**Figure 4 Derived smoke pollution profile for central Paris, 1893 using the Eiffel Tower data, together with an estimate of surface smoke pollution. Errors in the height determination are 10%.**